# Effective Hamiltonian based DNP Sequence Optimization

*Lorenzo Niccoli, Gian-Marco Camenisch, Matías Chávez and Matthias Ernst\**

Institute for Molecular Physical Sciences, ETH Zurich, CH-8093 Zurich, Switzerland

KEYWORDS: DNP; Floquet Theory; Pulse sequence optimization

ABSTRACT: Dynamic nuclear polarization (DNP) enhances the intensity of NMR signals by transferring polarization from electron spins to nuclei via microwave irradiation. Pulsed DNP methods offer more control on the spin dynamics than conventional continuous-wave approaches. Here, we report on-resonance and off-resonance DNP sequences optimized using effective Hamiltonians derived from continuous Floquet theory. Experiments at 80 K and 0.35 T using a sample of 5 mM Trityl OX063 in a glycerol-$d_8$/$D_2O$/$H_2O$ matrix (60:30:10, v/v/v) demonstrate that the optimized on-resonance sequence achieves 100 MHz electron offset bandwidth, while the off-resonance sequence cantered at an electron offset of 50 MHz can cover 20 MHz, with 25 MHz and 20 MHz of microwave power, respectively. These results demonstrate that continuous Floquet theory is a useful framework for the optimization of pulsed DNP sequences.

Dynamic nuclear polarization (DNP) is a powerful technique to enhance the NMR signals beyond the thermal equilibrium by transferring polarization from unpaired electrons to surrounding nuclei via suitable microwave (MW) irradiation[1–8]. Common applications of DNP rely on continuous-wave (CW) microwave irradiation to saturate certain transitions and have already extended NMR capabilities significantly by providing signal enhancement up to 3-4 orders of magnitude[9–12].

Besides the development of CW based DNP methodologies, there is increasing interest in pulsed DNP methods, because they allow better control of spin dynamics, in a similar fashion as in the transformation of NMR from CW to pulsed operation. Various DNP sequences have been developed, including NOVEL[13], off-resonance NOVEL[14], TOP-DNP[15], XiX-DNP[16], TPPM-DNP[17], BEAM[18] and frequency-swept DNP[19]. Currently many of the pulsed DNP sequences are limited to low magnetic fields due to the restricted availability of high-power amplifiers in the high GHz spectral range. Similar limitations in microwave power also affect CW-DNP experiments, particularly at high magnetic fields.

Modern pulsed EPR spectrometers use *arbitrary waveform generator* (AWGs) to generate the microwave pulses with precise control over phase, amplitude, and frequency[20]. The capabilities of modern AWGs open up new avenues to design pulse sequences for DNP. A recent example of such an approach is the PLATO sequence which can provide an excitation bandwidth of about 80 MHz across the electron resonance frequency[21].

Pulse sequence optimization is a wide area of research in magnetic resonance based on theorical frameworks such as the Average Hamiltonian Theory[22,23], Floquet theory[24–26], or single-spin vector effective Hamiltonian theory (SSV-EHT)[27–29]. However, these theoretical approaches usually work well for resonant and non-resonant terms in the effective Hamiltonian. The recently introduced Continuous Floquet theory[30,31] provides a more complete description of the spin dynamics by enabling also a good description in the near-resonance case which improves the description of recoupling sequence, e.g., MIRROR or symmetry-based C- and R- pulse schemes including resonance offsets[32].

In this manuscript, we present a method that allows the generation of optimized pulse sequences by employing standard gradient-based optimization algorithms to effective Hamiltonians derived from continuous Floquet theory. We demonstrate that this methodology enables the design of on- and off-resonance DNP sequences capable of achieving efficient polarization transfer. We show an on-resonance sequence with a bandwidth of 100 MHz and an off-resonance sequence that allows a 20 MHz bandwidth centred at an electron offset of 50 MHz, employing 25 MHz and 20 MHz of microwave amplitude respectively. We also compare this sequence optimization method with the one based on a figure-of-merit used for generating the PLATO sequence, by optimizing a sequence that aims for the same bandwidth which provided a similar enhancement but with about 21% lower microwave amplitude.

Continuous Floquet theory can be pictured as an extension of operator-based Floquet theory[32–34] by taking into account the finite length of pulse sequences. It requires a continuous frequency space description and not a discrete Fourier series representation. The finite length of pulse sequences leads to a broadening of resonance conditions, i.e., a convolution of the discrete Fourier series representation with a sinc function, enabling the correct description of near-resonance conditions and short non-periodic sequences. Calculation of the effective Hamiltonians is based on single-spin interaction-frame trajectories which makes the calculation of the effective Hamiltonian more efficient than exact multi-spin numerical simulations.

For a periodic time-dependent Hamiltonian of the form:

$$\widetilde{\mathcal{H}}(t) = \sum_n \widetilde{\mathcal{H}}^{(n)} e^{i\omega_n t} \tag{1}$$

the first order, closed-form effective Hamiltonian is given in continuous Floquet theory by:

$$\overline{\mathcal{H}}^{(1)} = \frac{1}{T}\widehat{\mathcal{H}}(0) = \sum_n \widetilde{\mathcal{H}}^{(n)} h_n^{(1)}(T) \tag{2}$$

with:

$$h_n^{(1)}(T) = \mathrm{sinc}\left(\frac{\omega_n T}{2}\right) \tag{3}$$

Here, we used a multi-index notation to write the Hamiltonian, where $\boldsymbol{n}$ is a tuple containing all the indices, $\boldsymbol{n} = (n, k, l)$ and $\omega_{\boldsymbol{n}} = n\omega_n + k\omega_k + l\omega_l$ is the weighted sum of all the characteristic frequencies. In Eqs. (1-3) $\widetilde{\mathcal{H}}^{(n)}$ are the Fourier coefficient and $T$ is the length of the sequence. It is important to realize that all Fourier coefficients potentially contribute to the first-order effective Hamiltonian, though their weight varies based on whether they are resonant ($\omega_{\boldsymbol{n}} = 0 \rightarrow \mathrm{sinc}(0) = 1$) or non-resonant ($\omega_{\boldsymbol{n}} \neq 0 \rightarrow |\mathrm{sinc}(0)| \leq 1$) terms. For the description of the pulsed DNP experiments Eq. (2) can be adapted considering all the relevant contributing frequencies in the interaction frame[35]:

$$h_{(n,k,l)}^{(1)}(T) = \mathrm{sinc}\left(\frac{(n\omega_{I,0} + k\omega_m + l\omega_\mathrm{eff})T}{2}\right) \tag{4}$$

where $\omega_m$ is the modulation frequency and $\omega_\mathrm{eff}$ is the effective field of the sequence while $\omega_{I,0}$ is the $^1$H Larmor frequency and $T$ is the length of the sequence defined by $T = (\sum_i \tau_i) \cdot n_r$ where $\tau_i$ is the pulse length of each pulse and $n_r$ is the number of times the sequence is repeated.

All the effective Hamiltonian calculations and optimization were conducted using MATLAB (The MathWorks Inc, Natick, MA, U.S.A). The effective Hamiltonian used for optimizations contains only the first-order term, as it has already been shown that this sufficient for accurately describing recoupling in NMR if resonance offsets are included in the interaction frame.[31] The optimization process involves minimizing the difference between the zero-quantum (ZQ) and double-quantum (DQ) and single quantum (SQ1, SQ2) components of the effective Hamiltonian. The optimization procedure starts from generating 6000 random initial pulse sequences and optimizing each of them with the gradient-based optimization algorithm *fmincon*. The generation of the starting sequences was performed using the *rng* MATLAB function, with the *twister* algorithm. The input parameters for the optimization are the bandwidth of the sequence in MHz and the pulse sequence parameters (the number of pulses, the pulse duration, and the maximum microwave power). The sequences were allowed to vary the amplitude in the [-1,1] range, i.e., corresponding to an amplitude of [0, 1] with phases of ±x. The effective Hamiltonian is only calculated for a single crystal orientation ($\beta = 45°$) as different crystallites have the same effective Hamiltonian except for a scaling factor.[36] For the calculation of the interaction-frame trajectory, the time step was set to 0.1 ns and the number of Fourier coefficient used to 30. A simple power distribution model was used to address microwave inhomogeneity[37], with normalized amplitude values of (1.05, 1.00, 0.95, 0.85) and corresponding weights of (0.1783, 0.3856, 0.2461, 0.1900).

The initial magnetization and the quantization axis of the effective Hamiltonian vary depending on the sequence type: in on-resonance sequences, it is aligned along the x-axis; in off-resonance sequences, it is aligned along the z-axis. At the end of the optimization procedure, the sequences with the largest difference between the ZQ and the sum of the ZZ, DQ, SQ1, SQ2 terms are selected. Our method is different from typical pulse-sequence optimization techniques[38] because it focuses on a cost function designed to maximize the components of the effective Hamiltonian that drive the polarization transfer and not the transferred polarization.

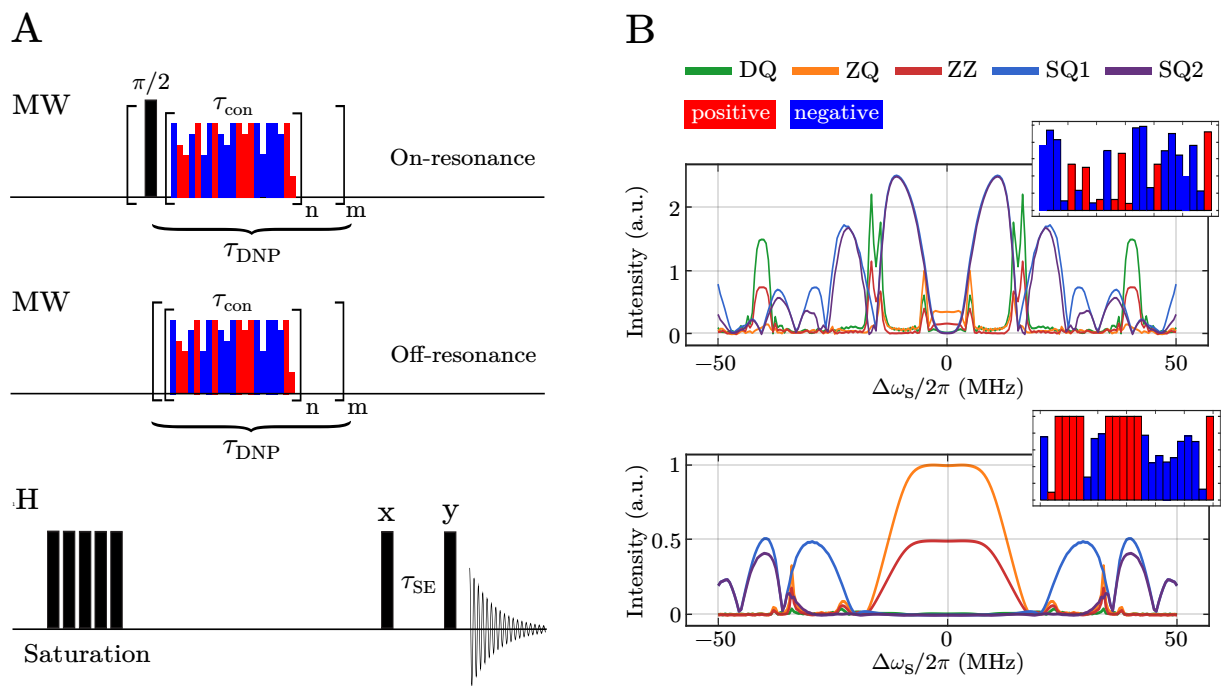

**Figure 1** (A) DNP pulse sequences used to evaluate the on-resonance (top left) and off-resonance sequences (middle left). (B) Representation of the optimization procedure. Starting from a random generated sequence (top), with random amplitude and randomly generated phases (±x, respectively - red and blue in the insets), the optimization procedure maximizes the ZQ term in the Hamiltonian that promote the polarization transfer (bottom).

The experimental evaluation of the optimized DNP sequences was conducted with a home-build X-band spectrometer analogous to the one described by Doll *et al*[20], at 80 K, on a sample of Trityl OX063 (5 mM) in glycerol-$d_8$/$D_2O$/$H_2O$ ("DNP juice", 60/30/10, v/v/v) using a protocol as described in Camenisch *et al*[35]. A schematic representation of the DNP experiments used for the acquisition of the on-resonance and the off-resonance sequences is shown in Fig. 1A. Each DNP sequence was started with a $^1$H saturation pulse train consisting of eleven 100° pulses to erase any proton thermal equilibrium polarization. Subsequently, each basic DNP block was repeated *n* times to give a total contact time $\tau_{con} = n\tau_m$. The total DNP experiment was then repeated *m* times to give a total build-up time $\tau_{DNP} = m\,\tau_{rep}$ where the shot repetition time $\tau_{rep}$ was 2 ms. For the on-resonance experiment a 90° pulse with phase +y was inserted before the DNP module. The number of repetitions, *m*, was 1000. The number of DNP cycles per repetition was optimized experimentally and a value of *n* = 3 was found to perform best for both the on- and off-resonance sequence. All the sequences were evaluated at their optimized power level. To account for the limited width of the microwave resonator mode and differences in non-linearity of the traveling wave tube amplifier at different frequencies, echo-detected nutation experiments were performed as described in Doll *et al*[39]. The hyperpolarized NMR signal was detected using a solid echo with pulse length of 2.5 µs spaced by 20 µs. For the solid echo an eight-step phase cycle was used with {x, x, y, y, -x, -x, -y, -y} for the first pulse and detection and {y, -y, x, -x, y, -y, x, -x} for the second pulse. The thermal-equilibrium reference experiment was recorded in the same way with the MW irradiation turned off and a delay of 180 s ($\approx 5 \cdot T_{1,n}$ where $T_{1,n}$ is the nuclear longitudinal relaxation time) was used between two consecutive scans. A total of 512 scans were recorded and accumulated for the reference experiment. The data has been processed as in Camenisch *et. al.*[35]. More details about the experimental setup, including the resonator profile and characterization of the TWT nonlinearity can be found in the SI.

We optimized both on-resonance and off-resonance sequences targeting the largest electron offset possible while aiming to use as little microwave power as possible. The on-resonance DNP sequences were designed to target a bandwidth of 100 MHz, using a maximum microwave power corresponding to a Rabi frequency of 25 MHz. The sequence is composed of 72 pulses, each with a duration of 5 ns, leading to a DNP sequence lasting 360 ns. The off-resonance sequence also consists of 72 pulses of 5 ns each and the centre was set at an electron offset of 50 MHz aiming for a bandwidth of 20 MHz, i.e., from an electron offset ranging from 40 MHz to 60 MHz, using a microwave power corresponding to a Rabi frequency of 20 MHz. Both sequences have a modulation frequency of $\frac{\omega_m}{2\pi} = 2.77$ MHz. The maximum rf-field amplitude for the sequences was chosen based on the resonator profile such that the Rabi frequency was accessible for the required offset range.

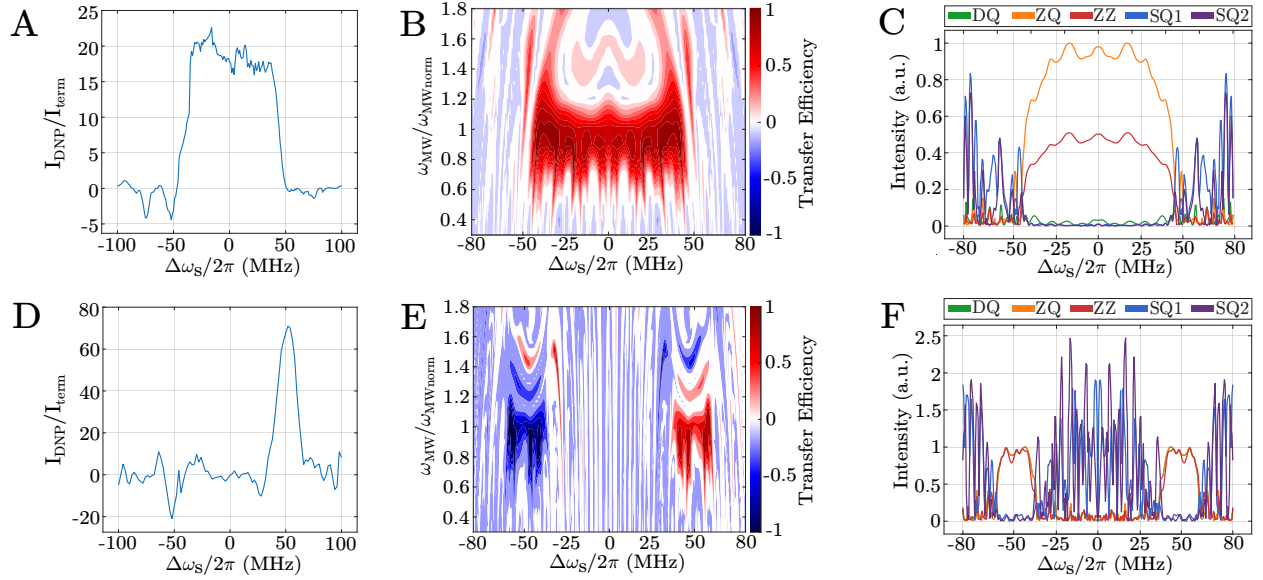

**Figure 2**. Subplot A, B, C refer to the on-resonance sequence while D, E, F to the off-resonance sequence. (A, D) Experimental DNP enhancement for the on-resonance (A) and off-resonance (D) DNP sequences, acquired with a repetition rate ($\tau_{\text{rep}}$) of 2 ms and a microwave power of 25 MHz and 20 MHz respectively. The values of the relative amplitude for each of the 72 pulses can be found in the SI. (B, E) Transfer efficiency as function of the normalized microwave amplitude (25 MHz and 20 MHz, respectively) and the electron frequency offset $\Delta\omega_S/2\pi$ for the two sequences. (C, F) Magnitude of the effective Hamiltonian terms as a function of the electron offset frequence $\Delta\omega_S/2\pi$, at a microwave amplitude of 25 MHz and 20 MHz, respectively. All the values are normalized to the maximum values of the ZQ term.

The experimental profiles for both sequences as a function of the electron offset $\frac{\Delta\omega_S}{2\pi}$ are shown in Figs. 2A-B, and have been acquired with a repetition rate ($\tau_{rep}$) of 2 ms i.e. $\tau_{DNP} = 2s$, resulting in peak enhancement values of about 23 and 70. Figures 2B and E show the transfer efficiency of the two sequences as a function of the microwave amplitude and the electron spin offset. For the on-resonance sequence there is good transfer efficiency within the optimized area and a satisfactory (±5%) tolerance to microwave inhomogeneity. Figures 2C and F show the magnitude of the relevant terms in the effective Hamiltonian as a function of the electron offset frequency. In both cases, the ZQ term dominates over the DQ, SQ1 and SQ2 terms, and these undesired terms are well suppressed across the selected bandwidth.

In order to compare our results to sequences described in the literature, we also optimized an on-resonance sequence that covered a bandwidth of 80 MHz but using a maximum MW amplitude of 25 MHz, and compared it with the recently published PLATO sequence (Figure 3A) that was optimized within the framework of the single-spin vector effective Hamiltonian theory (SSV-EHT)[27,28,29]. The PLATO sequence was acquired with a microwave power of 32 MHz, which is consistent with the value it was originally optimized for and is accessible over the relevant offset range on our experimental setup. The experimental comparison shows that our sequence can cover the same bandwidth with a similar enhancement value but using about 21% less microwave amplitude. In Fig 3B, we show the dependence of the transfer efficiency of the new sequence as a function of the microwave amplitudes and the electron offset frequency. The new sequence provides an optimal transfer across the offset frequency for microwave amplitudes in the range of 23-27 MHz, i.e., ±10% of the nominal value. The magnitude of the effective Hamiltonian term (Fig. 3C) clearly shows that our optimization procedure provides a good suppression of the unwanted DQ and SQ terms. In contrast, the PLATO sequence shows a contribution of the DQ term of around 7% compared to the ZQ term (see Fig. S5 of the SI). All the evaluated optimized DNP sequence, including PLATO, show a build-up time of about 6 seconds, which were recorded by incrementing the loop *m* in Fig. 1 at a given electron offset (0 MHz for the on-resonance sequence and 50 MHz for the off-resonance sequence) and keeping all the other parameter constant.

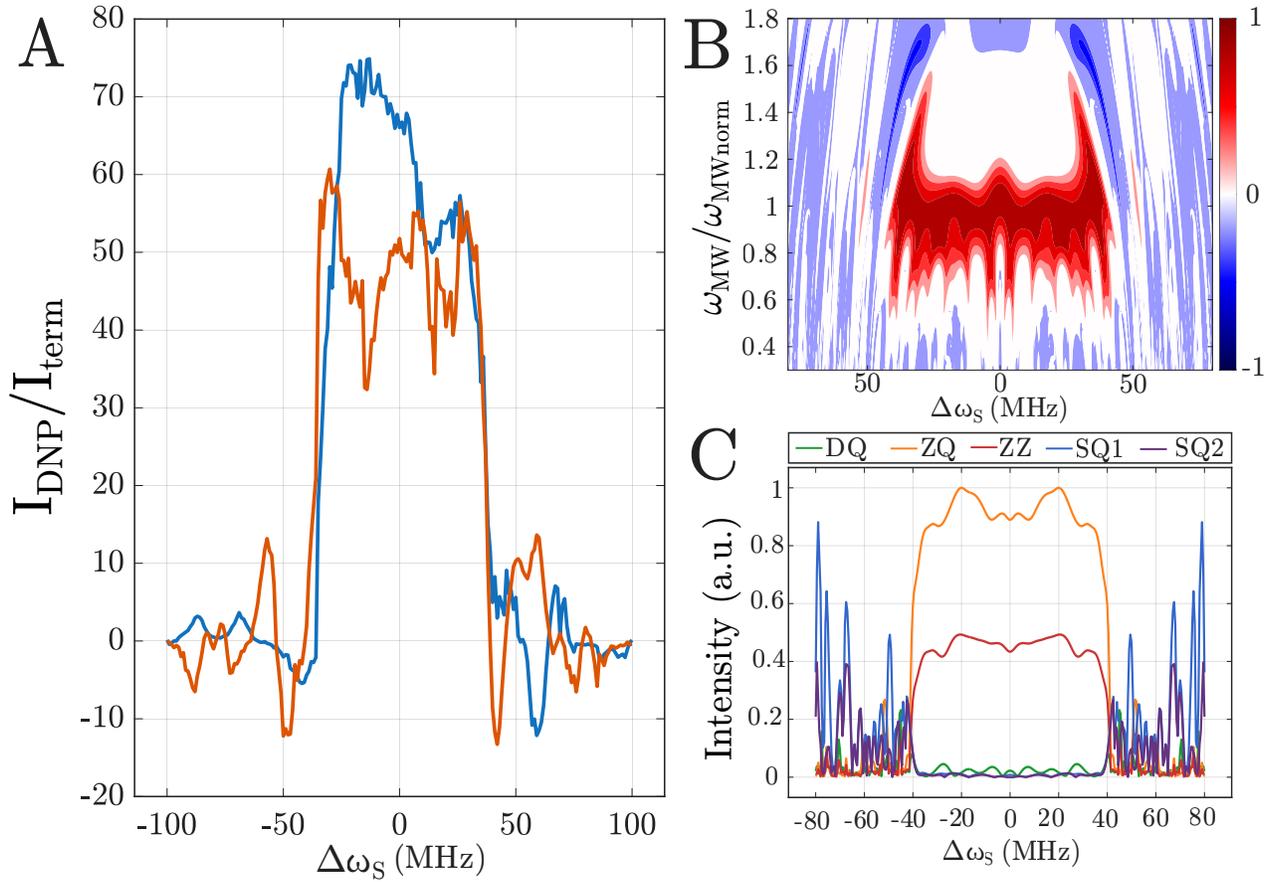

**Figure 3** (A) Comparison between the PLATO sequence and a new DNP sequence optimized to cover a bandwidth of 80 MHz. The PLATO sequence has been acquired with a repetition value of $n_r = 6$, while the new sequence used $n_r = 3$ leading to the same total length. For the PLATO sequence the microwave power was set to 32 MHz while for the new sequence was set to 25 MHz. Both sequences have a similar build-up time of ~6 s. (B) Transfer efficiency of the optimized DNP sequence as a function on the microwave amplitude (normalized to 25 MHz) and electron offset frequency $\Delta\omega_S/2\pi$. (C) Magnitude of the effective Hamiltonian terms (DQ, ZQ, ZZ, SQ1, SQ2) for the new DNP sequence as a function of the electron offset frequency $\Delta\omega_S/2\pi$. Similar plots for the PLATO sequence can be found in Fig. S5 of the SI

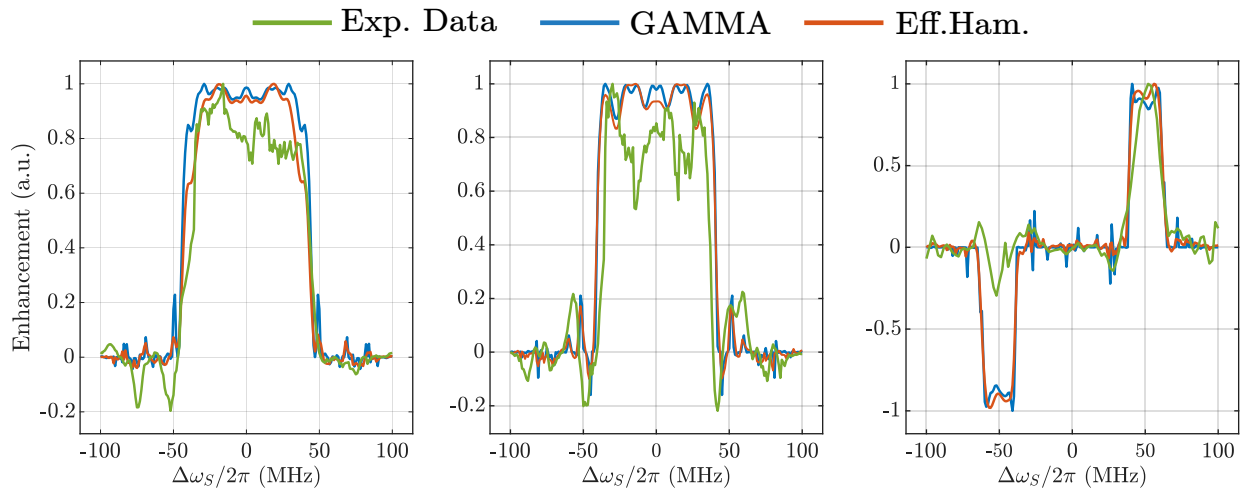

**Figure 4** Comparison of the DNP enhancement between experimental data (green), numerical spin-dynamics simulations (blue) and effective Hamiltonian based calculations (red). (A) On-resonance sequence optimized for a bandwidth of 100 MHz, (B) 80 MHz, and (C) off-resonance sequence centered a 50 MHz with a bandwidth of 20 MHz.

Besides the experimental characterization of the sequences, we have also characterized them by numerical simulations based on the calculated effective Hamiltonians and full numerical spin-dynamics simulations. Thus, effective Hamiltonian calculations have been compared to two-spin (1 electron, 1 proton) numerical simulations performed with the GAMMA[40] spin-simulation environment. The hyperfine coupling was calculated in ORCA 5[41,42] following the procedure described by Jeschke *et al*[43] and was set to 2.64 MHz (more details are reported in the SI). In Figure 4 we compare the experimental DNP profiles with profiles calculated both based on effective Hamiltonians and full numerical simulations. Each profile has been normalized to their respective maximum value to allow comparison. The comparison clearly shows that both the calculations based on the effective Hamiltonian and the simulations using full spin-dynamics simulations in a two-spin system are generally in good agreement with the experimental profiles, accurately reproducing the bandwidth. We note that the off-resonance sequence shows the biggest deviation from the experimental profile, especially in the range between -60 and -40 MHz. We attribute this discrepancy to the fact that the optimisation procedure targeted the area between +40 and +60 MHz, but the precise causes of this discrepancy are not yet understood. Other discrepancies between the simulated and measured results can be tentatively attributed to experimental imperfections, primarily phase transients and $B_1$ inhomogeneity, but their exact origin is still under evaluation. We also remark that power droop during the pulse sequence may play a significative role, as the power reduction over the time scale of the sequence could impact the transfer efficiency.

  Besides the sequences presented above, we further explored the capabilities of the optimization procedure by optimizing additional on-resonance and off-resonance pulse sequences targeting different bandwidths and using different numbers of pulses (48 or 96 each 5 ns long). For the on-resonance case, the optimized sequences span bandwidths of 80 or 120 MHz using microwave amplitudes of 25 or 30 MHz. The off-resonance sequences target bandwidths of 20 or 40 MHz, centered at electron offsets of 40 or 50 MHz. A comparison between the experimental results and the corresponding effective Hamiltonian and GAMMA simulations for these sequences is shown in Fig. S4. For all these additional sequences we find good agreement between effective

Hamiltonian and GAMMA simulations. In most cases (e.g., Fig. S4, sequences A and C), the targeted bandwidth was achieved, although the agreement between experiment and calculation was less satisfactory than for the sequences discussed in the main text. In other cases (e.g., Fig. S4, sequence D), the discrepancies were more pronounced. Regarding the off-resonance sequence, despite a good agreement between experiments and simulations (Fig S4, sequences E and F), we found lower enhancement values in the optimized region than for the off-resonance reported in Fig. 2. The underlying causes of all these discrepancies are still under investigation.

In summary, we have introduced on-resonance and off-resonance DNP pulse sequences at EPR X-band frequency (0.35 mT), optimized within the framework of continuous Floquet theory. Our approach is based on the optimization of the first-order effective Hamiltonian terms, i.e. DQ or ZQ, that promote the polarization transfer. Specifically, the optimized on-resonance sequence demonstrates a bandwidth of approximately 100 MHz using 25 MHz of microwave amplitude and the off-resonance sequence spans 20 MHz centered at an offset of 50 MHz. To benchmark our approach, we also optimized a sequence covering an 80 MHz bandwidth and compared it with the PLATO sequence designed for the same range. Our optimized sequence achieves a similar enhancement value and spans the same bandwidth efficiently while requiring about 21% less microwave amplitude, demonstrating improved transfer efficiency. While the total enhancement is higher for PLATO, reaching a maximum of about 72 compared our sequences still shows a comparable enhancement of about 60 (~20% lower). The first order effective Hamiltonian calculations were compared with numerical GAMMA simulations, and both were consistent with experiments, confirming the robustness of continuous Floquet theory to optimize pulsed DNP experiments. This proof-of-concept study demonstrates that the theoretical framework provided by continuous Floquet theory can be combined with gradient-based optimization algorithms, making it an additional tool for designing new DNP and NMR sequences. Nevertheless, we acknowledge cases in which significant discrepancies arise between the experimental data and calculations (Fig. S4) which are not fully understood. Starting from these observations, our current work focuses on developing additional sequences to further increase the achievable bandwidth and investigating the theoretical and experimental factors that influence the performance of the optimized DNP sequences.

## ASSOCIATED CONTENT

Supporting information. EPR/NMR spectroscopy: Resonator profile, TWT non-linearity, relative amplitude of the optimized pulse sequences, other optimized pulse sequences not discussed in the main text, effective Hamiltonian terms and transfer efficiency for the PLATO sequence, DFT calculations.

## AUTHOR INFORMATION

**Corresponding Author**

**Matthias Ernst** – *Institute for Molecular Physical Sciences, ETH Zurich, CH-8093 Zurich, Switzerland*.

email: maer@ethz.ch

**Authors**

**Lorenzo Niccoli** – *Institute for Molecular Physical Sciences, ETH Zurich, CH-8093 Zurich, Switzerland*

**Gian-Marco Camenisch** – *Institute for Molecular Physical Sciences, ETH Zurich, CH-8093 Zurich, Switzerland*

**Matías Chávez** – *Institute for Molecular Physical Sciences, ETH Zurich, CH-8093 Zurich, Switzerland*

**Notes**

Author contributions, according to CRediT guidelines: Conceptualization: LN, GC, ME; Data curation: LN, GC. Formal Analysis: LN, GC, ME. Funding Acquisition: ME, Investigation: LN, GC, ME; Methodology: LN, GC, MC, ME; Project Administration: ME. Resources: ME. Software: LN, GC, ME. Supervision: ME. Validation: LN, GC, ME. Visualization: LN. Writing

– Original Draft Preparation: LN, GC, ME. Writing – Review & Editing: LN, GC, MC, ME. The authors declare no competing financial interest. All authors have given approval to the final version of the manuscript.

ACKNOWLEDGMENTS

The authors gratefully acknowledge access to the pulsed EPR/DNP spectrometer by Prof. Gunnar Jeschke (ETH Zurich). ME acknowledges support from the Swiss National Science Foundation (grant no 200020 219375).

ABBREVIATIONS

AWG: arbitrary waveform generator. NMR: Nuclear Magnetic Resonance; DNP: Dynamic Nuclear Polarization; CW: Continuous Wave; FOM: Figure of Merit.

# Supporting Information: Effective Hamiltonian based DNP Sequence Optimization

*Lorenzo Niccoli, Gian-Marco Camenisch, Matías Chávez and Matthias Ernst\**

Institute for Molecular Physical Sciences, ETH Zürich, CH–8093 Zürich, Switzerland

**Table of Contents**

# A. Pulsed DNP experiments

In this section we report the resonator profile and the measure of the non–linearity of the travelling wave tube (TWT) amplifier of the spectrometer used to characterize all the pulsed DNP sequences reported in the main text (Fig. S1-S2). The experimental protocols used for these measurements are analogous to the ones described in Camenisch et al[1].

## Resonator Profile

The experiment to record a resonator profile is a three-pulse experiment. The first pulse serves as a nutation pulse and is incremented from 0 to 128 ns in steps of 2 ns at maximum power (digital scale = 1). Electron spin magnetization after a delay T~$5T_{2,e}$ is observed with a two-pulse Hahn echo experiment. The experiment is measured for different electron offsets with respect to the center of the resonator. The external magnetic field is swept to be on resonant with the mw frequency. The resulting resonator profile indicates the largest $B_1$ field (or strongest Rabi frequency) that can be obtained at a certain electron offset.

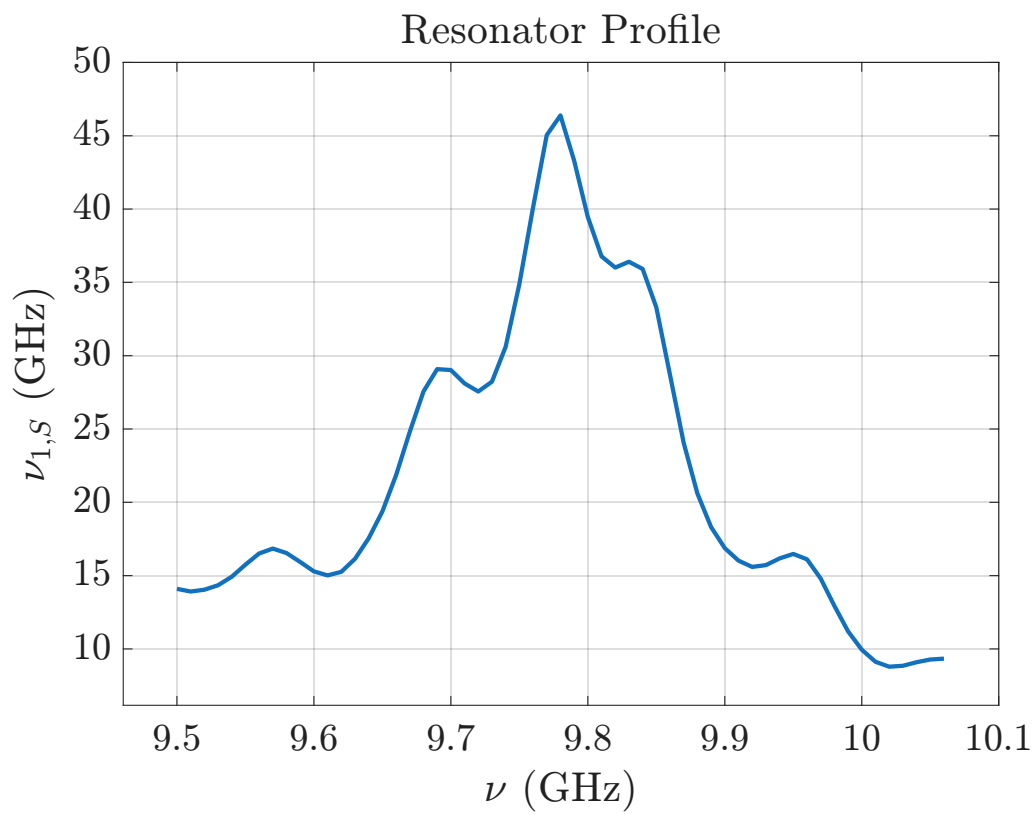

**Figure S1** Resonator profile for the trityl radical. This profile indicates the largest $B_1$ field that can be obtained at a certain offset. The center of the resonator is at $\nu = 9.78$ GHz.

# Non–linearity of the Travelling Wave Tube (TWT) Amplifier

The experiment to record the non-linearity of the TWT amplifier is a three-pulse experiment. The first pulse serves as a nutation pulse and is incremented from 0 to 512 ns in steps of 2 ns at various digital scale. Electron spin magnetization after a delay a delay T~$5T_{2,e}$ is then observed with a two-pulse Hahn echo experiment. The experiment is measured at the center of the resonator ~ 9.78 GHz. Fitting for both the dependence of Rabi frequency on digital scale and the dependence of digital scale on required Rabi frequency of the TWT non-linearity curve by polynomials of fourth order allows the mapping between the digital scale and the Rabi frequency. The TWT non-linearity curve together with the resonator profile is used to compensate the limited width of the microwave resonator mode and differences in non-linearity of the TWT during the acquisition of a DNP profile.

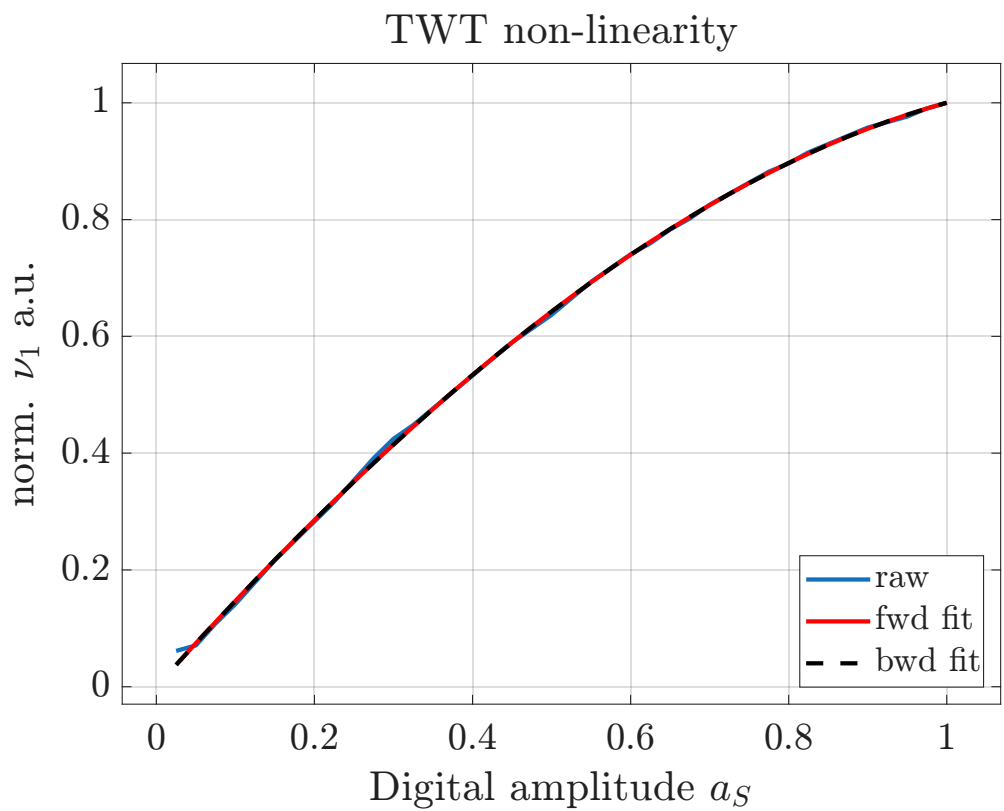

**Figure S2** Non-linearity of the TWT amplifier for the trityl radical. The TWT non-linearity curve together with the resonator profile is used to compensate the limited width of the microwave resonator mode and differences in non-linearity of the TWT during the acquisition of a DNP profile.

# B. Pulsed DNP sequences

## Sequences with 72 pulses

In this section we report the relative amplitude of all the optimized sequences discussed in the main text. The offset for the off-resonance sequence is +50 MHz. A graphical representation of each pulse sequence in shown in Fig. S3.

### On-resonance DNP sequence

**Bandwidth = 100 MHz, microwave power = 25 MHz**

[0.2799  0.9119  1.0000  1.0000 −0.5883 −1.0000  0.0340 −1.0000 −1.0000 −1.0000 −1.0000  1.0000  1.0000 −0.2218 −1.0000 −1.0000 −1.0000 −0.6442  1.0000  0.7882 −0.1632  1.0000  1.0000  0.8996 −1.0000  0.0042 −0.9758 −1.0000 −1.0000 −1.0000 −1.0000  1.0000  1.0000  0.2030 −1.0000  0.0389 −1.0000 −1.0000 −0.5097  1.0000  1.0000  0.9171 −0.6976 −0.5858  1.0000  1.0000  1.0000  0.6177 −1.0000  0.7784  1.0000  0.1306  0.9129  1.0000 −1.0000 −1.0000 −1.0000 −0.9784 −0.5475  0.5806  1.0000  1.0000 −0.1561 −0.7739  1.0000  1.0000  1.0000  1.0000 −0.5312 −1.000  0.9177  1.0000]

**Bandwidth = 80 MHz, microwave power = 25 MHz**

[ −1.0000 −1.0000 −1.0000  0.8048  1.0000  0.6833  0.8304  1.0000  0.2942 −1.0000 −1.0000 −0.4535 −1.0000 −1.0000 −1.0000 −1.0000  1.0000  1.0000  0.5833 −1.0000 −0.8229 −0.7501 −1.0000 −1.0000  1.0000  1.0000  1.0000 −0.4981 −1.0000  1.0000  1.0000  1.0000  0.9914 −1.0000  0.5821  1.0000  0.4121  1.0000  1.0000 −1.0000 −1.0000 −1.0000 −0.8430 −0.5235  0.0848  1.0000  1.0000  0.7286 −1.0000 −0.5017  1.0000  1.0000  1.0000  1.0000 −1.0000 −0.1049  1.0000  1.0000  1.0000  1.0000  1.0000 −1.0000 −1.0000 −0.9327 −0.1575 −0.8397 −0.5034 −1.0000  0.5989  1.0000  1.0000 −1.0000]

## Off-resonance DNP sequence

**Bandwidth = 20 MHz, microwave power = 20 MHz**

[  1.0000  −0.8886  −1.0000   1.0000   0.7763  −1.0000  −1.0000   1.0000   0.5991  −0.2579
 1.0000   1.0000   1.0000  −1.0000  −1.0000   0.5101   1.0000   0.0544   0.3526  −0.3569  −1.0000
 0.4502   1.0000  −1.0000  −1.0000   1.0000   1.0000  −1.0000  −1.0000   1.0000   1.0000  −1.0000
 0.6908  −1.0000  −1.0000   1.0000   1.0000  −1.0000   0.3166   0.2955  −1.0000  −1.0000   1.0000
 1.0000  −1.0000  −1.0000   1.0000   1.0000   1.0000  −1.0000   1.0000   1.0000  −1.0000  −1.0000
 1.0000  −1.0000  −1.0000   1.0000  −0.4263  −1.0000  −0.0379   1.0000   1.0000  −1.0000  −1.0000
 1.0000   1.0000  −0.8975   0.8810  −1.0000  −1.0000   1.0000]

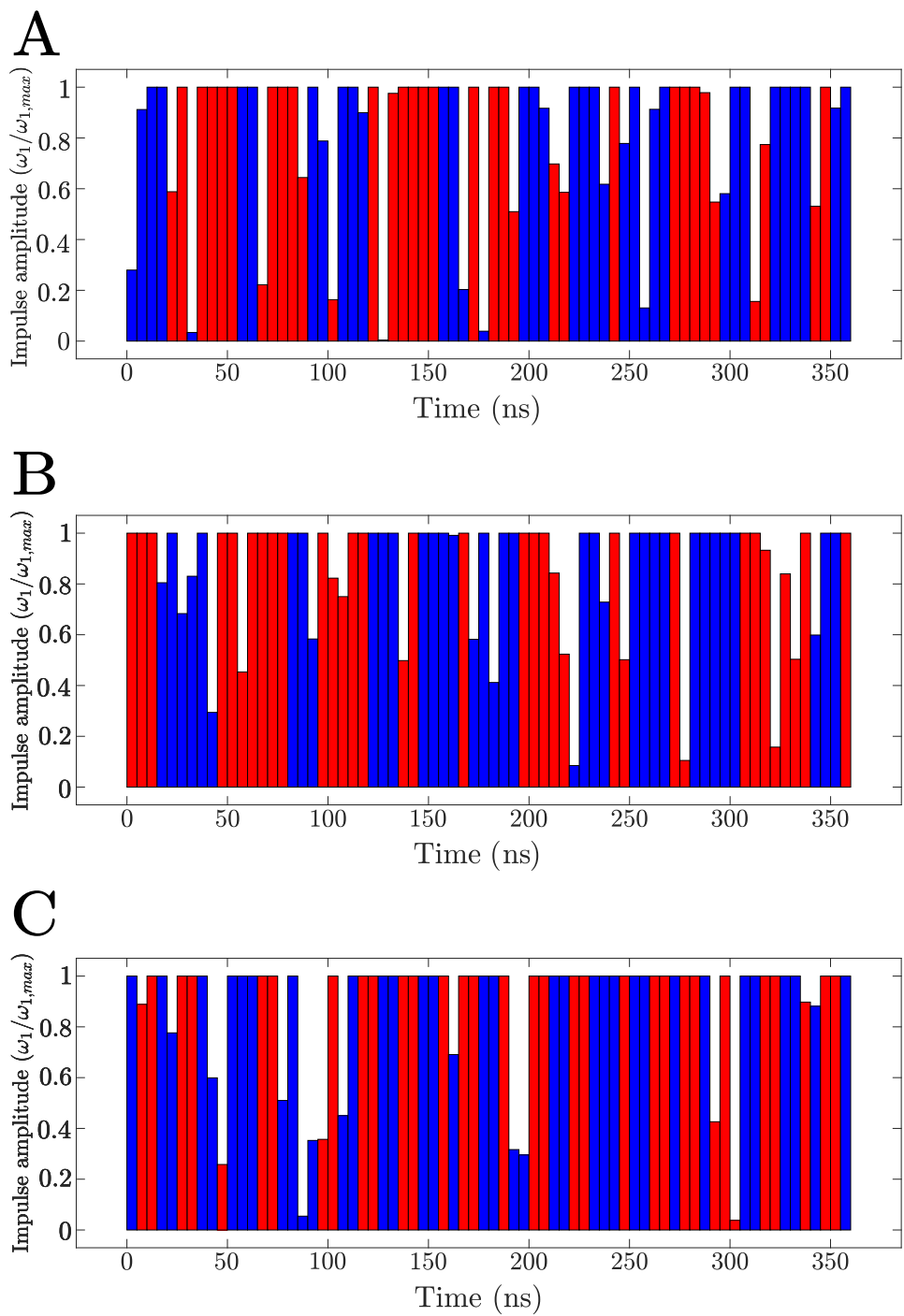

**Figure S3.** Pulse sequence diagrams: (A) On-resonance, 100 MHz bandwidth, 25 MHz amplitude; (B) On-resonance, 80-MHz bandwidth, 25 MHz amplitude; (C) Off-resonance (50 MHz center), 20 MHz bandwidth, 20 MHz amplitude. Blue and red denote positive and negative amplitudes.

# Other pulsed DNP sequences

In this section, we report additional optimized DNP sequences that are not discussed in the main text. These include both on-resonance and off-resonance sequences consisting of 48 or 96 pulses each with 5 ns length, covering electron offset bandwidths of 80 MHz or 120 MHz for the on-resonance case, and bandwidths of 20 MHz or 40 MHz centered at 40 or 50 MHz for the off-resonance case. The parameters for all optimized sequences are summarized in Table S1, and the sequences are labeled A–F for convenience. The corresponding experimental data, acquired as described in the main text, are compared with effective Hamiltonian calculations and GAMMA simulations in Figure S4. The repetition number was set equal to 4 for the sequences with 48 pulses and to 2 for the sequences with 96 pulses. The relative amplitudes of all these sequences are reported later in this section.

**Table S1**. Sequence type, number of pulses, microwave power and target bandwidth for the DNP sequences (A-F) presented in this section.

| Label | Sequence type | N° pulses | Microwave Power (MHz) | Bandwidth (MHz) |
|---|---|---|---|---|
| A | On-resonance | 48 | 25 | 80 |
| B | On-resonance | 96 | 25 | 120 |
| C | On-resonance | 48 | 30 | 80 |
| D | On-resonance | 96 | 30 | 120 |
| E | Off-resonance | 48 | 15 | 20 (center = 40 MHz) |
| F | Off-resonance | 96 | 15 | 40 (center = -50 MHz) |

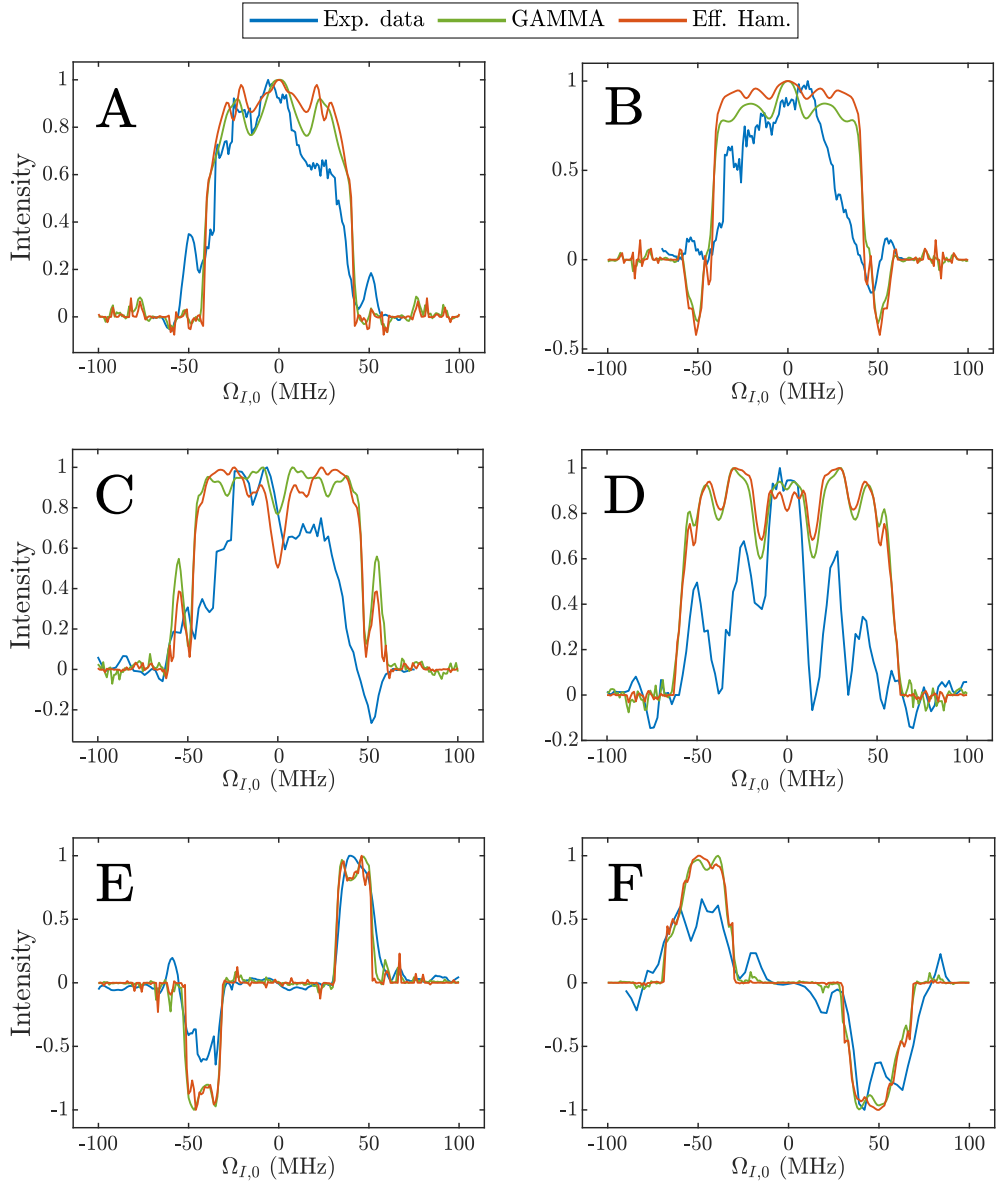

**Figure S4**. Comparison of experimental data with effective Hamiltonian and GAMMA calculations for different optimized pulse sequences. Sequences A and C were acquired with 25 MHz microwave power, B and D with 30 MHz, and E and F with 15 MHz. On-resonance sequences: (A) 48 pulses, bandwidth: 80 MHz; (B) 96 pulses, bandwidth: 120 MHz; (C) 48 pulses, bandwidth: 80 MHz; (D) 96 pulses, bandwidth: 120 MHz. Off-resonance sequences: (E) 48 pulses, bandwidth: 20 MHz centered at 40 MHz; (F) 96 pulses, bandwidth: 40 MHz centered at 50 MHz. Each profile has been normalized to their respective maximum value to allow the comparison.

**Sequence A**

**N° Pulses = 48, Bandwidth = 80 MHz, Microwave Power = 25 MHz**

[ -0.2888 -1.0000  0.6968  1.0000  1.0000  1.0000 -0.6580  0.9052  0.5621  1.0000  1.0000
-1.0000 -1.0000 -1.0000 -1.0000 -0.3123 -0.0464  1.0000  1.0000  0.7573 -1.0000 -0.6967
1.0000  1.0000  1.0000  1.0000 -1.0000  0.5312  1.0000  1.0000  0.7598  1.0000  1.0000
-1.0000 -1.0000 -1.0000  0.3287 -1.0000 -1.0000 -0.3524  1.0000  1.0000  0.9471 -1.0000
-1.0000 -1.0000 -1.0000  0.3741 ]

**Sequence B**

**N° Pulses = 96, Bandwidth = 120 MHz, Microwave Power = 25 MHz**

[ -0.9997  1.0000  1.0000  1.0000 -0.4583  0.9999 -0.5605 -0.9999  1.0000  1.0000  1.0000
0.4129 -0.9985  0.5609  0.9999  0.6287 -1.0000 -1.0000 -0.7696 -0.3387 -0.5094  1.0000
1.0000 -0.3268 -1.0000 -1.0000 -1.0000 -0.1542 -1.0000 -1.0000 -1.0000  1.0000  1.0000
-0.4811  0.5559  1.0000  0.5627 -1.0000 -1.0000 -1.0000 -0.2017  1.0000  0.2772 -1.0000
-1.0000 -0.9717  0.7483 -1.0000 -1.0000  1.0000  1.0000  1.0000 -0.0913  1.0000 -0.2923
-1.0000  1.0000  1.0000  1.0000  0.3393 -0.9999  0.3198  1.0000  0.2831 -1.0000 -1.0000
-0.3647 -0.7959  0.1266  1.0000  1.0000 -0.4888 -1.0000 -1.0000 -1.0000 -0.0839 -1.0000
-1.0000 -0.7032  1.0000  0.9720 -0.5770  1.0000  1.0000  0.0723 -0.8969 -1.0000 -1.0000
0.1382  1.0000 -0.0281 -0.9884 -0.7306 -0.9999  0.3825 -0.3546 ]

## Sequence C

**N° Pulses = 48, Bandwidth = 80 MHz, Microwave Power = 30 MHz**

[  0.0536  -1.0000  -1.0000  -1.0000  -0.8482   0.8101   1.0000   0.9027  -0.6312  -0.2786
-0.9251   0.0870  -0.8777  -1.0000  -1.0000   1.0000   1.0000   0.4270   1.0000   1.0000
-0.0302  -1.0000   1.0000   1.0000   1.0000   1.0000  -0.7618  -0.3275   0.8531   0.9239
 0.3897  -0.5858  -0.8759  -1.0000  -1.0000  -1.0000   1.0000   1.0000   0.5500   0.4936
 0.0071  -1.0000   1.0000   1.0000   1.0000   0.4190  -1.0000   1.0000]

## Sequence D

**N° Pulses = 96, Bandwidth = 120 MHz, Microwave Power = 30 MHz**

[ -0.3448  -1.0000  -0.4251   1.0000   1.0000   1.0000   1.0000  -1.0000  -1.0000   0.9999   0.5869
-0.3181   1.0000   1.0000   1.0000  -0.6460   0.0252   0.4508   0.1485   0.4072  -1.0000  -1.0000
-0.9846   1.0000  -0.1442   0.7468   1.0000   1.0000  -1.0000  -0.0317   1.0000   1.0000   1.0000
-0.8755  -1.0000   0.8315   1.0000  -0.1497   1.0000   1.0000   0.7510  -1.0000  -0.6639   0.4474
-1.0000  -1.0000  -1.0000   0.9999   0.5893  -1.0000  -1.0000  -0.9999  -1.0000   0.6244   1.0000
-0.3509  -0.1631   1.0000   1.0000  -0.9200  -0.6451  -0.9996  -0.1525  -0.6436  -0.9999  -1.0000
 0.7728   1.0000   0.8722  -1.0000  -0.1966  -0.1675  -1.0000  -1.0000   1.0000   1.0000   0.2215
 1.0000   0.9999   0.2195   0.8148  -1.0000  -1.0000   0.4682   0.6243   1.0000   1.0000   1.0000
-0.6603  -0.4285   0.9680   1.0000  -0.4778  -1.0000  -1.0000   0.9999 ]

**Sequence E**

**N° Pulses = 48, Bandwidth = 20 MHz (centre = 40 MHz), Microwave Power = 15 MHz**

[ -1.0000  1.0000  1.0000  1.0000  1.0000 -1.0000 -1.0000 -1.0000  1.0000  1.0000 -1.0000 -1.0000  0.3870  1.0000  1.0000 -1.0000 -1.0000  1.0000 -1.0000 -1.0000  1.0000  1.0000 -1.0000 -0.7340  1.0000  1.0000  1.0000 -1.0000 -1.0000  1.0000  1.0000  1.0000 -0.0950 -1.0000 -1.0000  1.0000  1.0000 -1.0000 -1.0000  1.0000  1.0000 -0.8733 -1.0000  0.8681  0.7675  0.8575 -1.0000 -1.0000 ]

**Sequence F**

**N° Pulses = 96, Bandwidth = 40 MHz (centre = -50 MHz), Microwave Power = 15 MHz**

[-1.0000  1.0000 -0.4483 -1.0000  0.7674  0.7882 -1.0000 -0.7308  0.3295  1.0000  0.8511 -1.0000 -1.0000  0.6118  1.0000 -0.1929  1.0000  1.0000 -1.0000 -1.0000  1.0000 -0.1193 -0.1778  1.0000 -1.0000 -1.0000  1.0000  1.0000 -0.8296 -1.0000  1.0000  1.0000 -1.0000 -1.0000  1.0000 -0.0648  1.0000  1.0000 -1.0000 -1.0000  1.0000  0.6921  0.2611 -0.6332 -1.0000  1.0000  0.8796 -1.0000 -0.9336  1.0000 -0.8427 -0.7389  1.0000  1.0000  0.0141 -1.0000 -0.2760  1.0000  1.0000 -1.0000 -1.0000 -1.0000  0.1884 -0.6529  1.0000  1.0000 -1.0000 -1.0000  0.7279 -1.0000 -0.5143  1.0000 -1.0000 -1.0000  1.0000  1.0000 -1.0000 -0.9277  1.0000  1.0000 -1.0000 -1.0000  0.4094  1.0000  0.3373  1.0000 -1.0000 -1.0000  1.0000  1.0000 -0.5328  1.0000 -1.0000  1.0000  1.0000 -1.0000]

# C. Effective Hamiltonian terms and transfer efficiency for the PLATO sequence

In Figure S5 we report the calculation for the PLATO sequence[2] of the effective Hamiltonian terms as a function of the electron frequency offset, together with the corresponding transfer efficiency evaluated as a function of the normalized microwave amplitude (32 MHz) and the electron frequency offset $\Delta\omega_S/2\pi$. As in the main text, all effective Hamiltonian terms are normalized to the maximum value of the ZQ term.

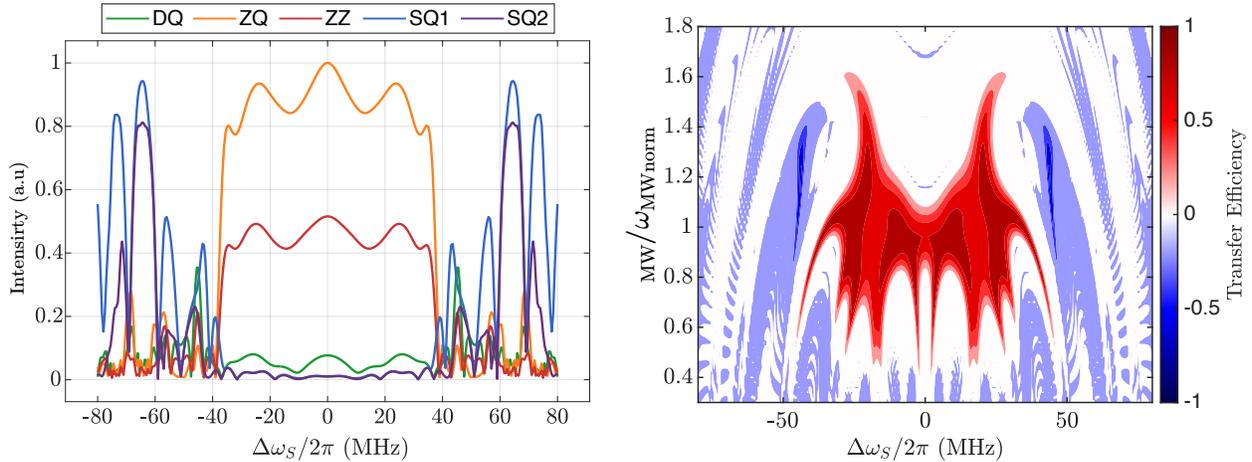

**Figure S5** (Left) Effective Hamiltonian terms (DQ, ZQ, ZZ, SQ1, SQ2) in function of the electron frequency offset. (Right) Transfer efficiency as a function of the electron frequency offset and the microwave amplitude.

# D. DFT Calculation on the Trityl OX063 radical

The DFT calculation on Trityl OX063 have been performed with ORCA 5[3,4]. The initial geometry was generated with Avogadro[5] and optimized using the B3LYP[6,7] functional with D3BJ[8,9] dispersion corrections, the def2-SVP basis set, and TightSCF options.

Hyperfine couplings were then computed on the optimized geometry using the B3LYP functional, the EPR-III basis set[10], and the D3BJ and TightSCF options. For sulfur atoms was set the def2-TZVPPD[11,12] basis set.

The most intense hyperfine coupling (2.64 MHz) was used as the reference value for the GAMMA[13] calculations presented in the main text.